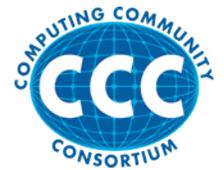

# From Data to Knowledge to Action:
# A Global Enabler for the 21st Century

Eric Horvitz  
Microsoft Research

Tom Mitchell  
Carnegie Mellon University

**Unprecedented advances**

A confluence of advances in the computer and mathematical sciences has unleashed unprecedented capabilities for enabling true *evidence-based decision making*.  These capabilities are making possible the large-scale capture of data and the transformation of that data into *insights* and *recommendations* in support of decisions about challenging problems in science, society, and government.

Key advances include jumps in the *availability of rich streams of data*, precipitous drops in the cost of *storing and retrieving massive amounts of data*, exponential increases in *computing power and memory*, and jumps in the prowess of *methods for performing machine learning and reasoning*.  These advances have come together to create an inflection point in our ability to harness large amounts of data for generating insights and guiding decision making.

The shift of commerce, science, education, art, and entertainment to the web makes available unprecedented quantities of structured and unstructured databases about human activities – much of it available to anyone who wishes to mine it for insights.  Logs of interactions are regularly captured by web services such as search engines and online shopping sites and with specialized point-of-sale terminals.  Outside the web, organizations regularly capture transactions and operational data on enterprise databases.  Sensor networks are sprouting in multiple venues in support of a wide spectrum of applications, from sets of traffic flow sensors in highways, to meshes that track moisture in vineyards.  Beyond fixed sensors, dynamic sensor networks are being created by constellations of connected mobile devices (e.g., GPS data).

In the sciences, new evidential paradigms and sensing technologies are making available great quantities of data, via use of fundamentally new kinds of low-cost sensors (e.g., genomic microarrays) or through viewers that provide unprecedented scope and resolution.  Consider for example the Sloan Digital Sky Survey (SDSS), an astronomical survey kicked off in 2000 to capture 100 million stellar objects, and now in a third phase of study, aimed at capturing finer structural details about our galaxy and more distant bodies.  Today, 200 GB of new high-resolution optical data are being captured each evening by charge-coupled devices (CCDs) attached to telescopes.  The data pose a huge opportunity for data-centric analyses.

---

To date, we have only scratched the surface of the potential for learning from these large-scale data sets. Opportunities abound for tapping our new capabilities more broadly to provide insights to decision makers and to enhance the quality of their actions and policies.

**Optics onto the Unknown**

A constellation of algorithmic tools developed within computer science can transform case libraries – typically impenetrable rows and columns of raw data – into insights, in the form of rich visualizations and recommendations for action. The tools include approaches for discovering and visualizing important dependencies, influences, and relationships amidst a sea of complex interactions, as well as methods for constructing and using predictive models. Predictive models predict things that we cannot see directly from the things that we can observe with ease.

Predictive models constructed from historical data have already provided an "optics" onto such hidden states as the potential illnesses lurking behind a patient's symptoms, the durations of traffic jams in a city, the best way to translate a sentence from English into Arabic, the informational goals lurking behind a short query input to a search engine, and the influences of patterns of genes on the risk of acquiring a disease.

**From Data to Decisions**

Great strides have been made recently in developing *machine learning algorithms* that automatically discover predictive models from historical data. These are now used routinely to predict which credit card transactions are likely to be fraudulent, which emails are likely to be spam, and which handwritten letters are intended for which zipcode. These predictive models often suggest several alternative outcomes when the prediction is inherently uncertain, along with probabilities for each. These probabilities can then be used to identify ideal *decisions*. Such decisions include actions in the world, as well as decisions about when it is best to collect more information about a situation before acting, considering the costs and time delays associated with collecting more information to enhance a decision. Great progress has been made in assessing the value of collecting additional information before taking action. Recommendations for action are generated via a pipeline of data → predictive models → decision analysis. The likelihoods output by the predictive models lay at the heart of decision analyses that weigh the costs and benefits of different actions. We describe below (and, in further detail, in the companion reports) several key areas of national priority in which applying this data → predictive models → decision analysis pipeline is critical to ensuring our nation's prosperity and wellbeing.

**Enabling Evidence-Based Healthcare**

Critically valuable data continues to fall on the floor in healthcare, where paper records and manual procedures for information capture still dominate the landscape. There is tremendous upside to investing in data collection and machine learning in healthcare. Predictive models can be harnessed to simultaneously enhance the quality of healthcare and lower its costs. Opportunities for machine learning in healthcare include the construction of predictive diagnostic models that are pressed into service when performing automated triage, inter-specialty referral, and diagnosis of patients presenting with sets of complaints, signs and symptoms. Broader applications include the use of predictive models to optimize the management of chronic diseases, a chief source of healthcare costs and the root of poor quality of life for our aging population. In one approach, predictive models can inform how scarce resources should be applied for maximal benefit. As an example, a recent study of 300,000 electronic records of visits to emergency departments in the Washington, DC, area highlights the value of machine learning to lower costs by minimizing re-hospitalizations within 30 days through selective investment in post-discharge support where it will be most useful. Machine learning will also provide a foundation for clinical discovery that promises to identify associations between gene composition and risks of illness, disease progression, and the efficacy of pharmacological agents, and the broader hope of identifying interventions and cures. Within the next 12-36 months, people will be able to have their entire genome sequenced for less than $1000 – a near future that one expert referred to as a "a 10-mile-wide asteroid heading toward us." There are great opportunities to learn from the massive quantity of data that will likely become available, and multiple issues, from storage to privacy to methods for learning from large amounts of genomic data, must be confronted in the near term.

**Enabling the New Biology**

Over the last several decades, new experimental techniques coupled with advanced data analytics strategies have utterly transformed the way we study biology. Pursuit of this field has become an increasingly data-centric endeavor, now comprising iterative stages of computation and experiment. Today, genomes of plants and animals are typically captured with gigabytes of data, and machine learning is used to identify networks and modules at the core of the operation of cells. As our knowledge has grown, so, too, has our appreciation for the complexity of biological systems. For example, at a minimum, there is simply no way to grasp epigenetic processes (i.e., studies of the control of RNA transcription, splicing machinery, and the selective expression of proteins, including the combinatorics therein) without computation. Indeed, biological function arises from the complex interplay of individual components (genes, proteins, metabolites), and the emerging science of systems biology – which strives to integrate biological structures and create predictive models representing holistic functions and behaviors – offers the potential to radically alter how we understand disease, develop new drugs and interventions (taking into account system-wide effects to identify potential side effects early on), and engineer organisms to synthesize important byproducts such as biofuels. Biological data capture and analytical requirements quickly scale to multiple terabytes of information – capturing information at each time slice of periods of study – and computation therefore sits at the leading edge of discovery in biology; it will continue to be front and center in building insights and moving forward with models and understandings amidst complexity.

**Enabling 21st Century Discovery in Science and Engineering**

Computer science is more generally transforming all of the major sciences into eSciences – reshaping the way science is carried out in the 21st century. Scientific efforts have long relied on extracting insights from measurements, models, and simulations. However, data-centric methodologies have grown in importance in the sciences with the growing flood of data, and concomitant rise in power of computing resources.  Data has grown by orders of magnitude over the last 20 years with the use of such tools as automated sky surveys in astronomy, gene sequencing in biology, web logs of behavioral data in sociology, and the deployment of sensor networks in agriculture and the environment.  New information-centric subdisciplines, such as *bioinformatics*, *astroinformatics*, and *matinformatics* (data-centric computing in materials science) have arisen. Computational tools critically enable discovery and inference with methods for simulating phenomena, visualizing processes, extracting meaning from massive data sets – and directing the collection of new data.  Computational models for learning and inference also show great promise in their ability to provide guidance to scientists on the overall hypothetico-deductive cycle of discovery, where an initial predictive model learned from data guides the iterative collection of new data, followed by model revision and new inferences that once again point the way to new data to collect.  Advances in machine learning have multiple touch points with the process of science.  The methods even cut to the core goals of scientific inference with algorithms that can elucidate or rule out the existence of causality in processes.  The melding of new data resources with methods of simulation, visualization, and machine learning will enable scientists to pursue insights and comprehension at a faster pace than ever before in many key areas of scientific discovery.

**Enabling a Revolution in Transportation**

Opportunities abound for enhancing the efficiency and reliability of multiple modalities of our transportation system with data-centric approaches.  Recent efforts with machine learning on large-scale data sets about flows of traffic demonstrate how we can use predictive models to predict future road flows – and flows on roads that are not sensed in real time.  Such predictions can be harnessed for generating context-sensitive directions, real-time routing, and load balancing.  Predictive models for traffic flows in greater city regions have leveraged years of heterogeneous data, including in-road sensors, GPS devices in vehicles (mass transit and crowd-sourced from volunteers), information about road topology, accidents, weather, and major events such as sporting events.  This work has led to fielded services that provide flow prediction and directions in major cities of the United States.  Other opportunity areas in transportation include integrating predictive models with fluid dynamics and queue-theoretic models to build tools that help engineers to understand the value of new roadways and road upgrades.  We can also apply simulations to understanding the real-world implications of different high-occupancy vehicle (HOV) and high-occupancy toll (HOT) policies and provide insights about the demand and flows on roads expected with different designs for dynamic tolling systems.  In another area, meshing predictive models with automated planning systems can help to shift people from private vehicles to shared transportation solutions.  For example, automated planning systems can assemble *multimodal transportation plans* that mix public and private transportation, offering people more flexible, end-to-end transportation alternatives.  Predictive models combined with optimization can also be employed to mediate and optimize the operation and ease of participating in ridesharing systems.  Data collection and learning models for driving, combined with advances in real-time sensing, will also play an important role in building safer cars that employ collision warning and avoidance systems.  Such systems promise to reduce the unacceptably large number of deaths on U.S. roadways each year (more than 40,000 deaths and even higher numbers of disabling injuries per year).  Methods for learning automated driving competencies from data will additionally be crucial in the development of autonomous vehicles that drive without human intervention.  Beyond enhancing the safety of transport in private vehicles and enabling greater densities of vehicles on freeways, the development of autonomous vehicles will enable the creation of large-scale public microtransit systems that flexibly transport people from point to point within city regions, continuing to execute prediction and optimization for maximizing their availability and efficacy.  The latter potential shift in the way we do most of our travel has deep implications for energy and the environment.

**Enabling Advanced Intelligence and Decision-Making for America's Security**

Data-centric learning and modeling have an important and promising role to play at multiple levels in America's security.  Intelligence, surveillance, and reconnaissance (ISR) poses challenges that are well suited for data-centric computational analyses.  Statistical methods for fusing information have undoubtedly been used for evidential analysis by government agencies engaged on security.  However, late-breaking computing research on such topics as active learning, learning from network features, and the use of reliability indicators, can enhance our ability to effectively piece together data from heterogeneous sources and sensors.  Machine learning and reasoning can be applied in both exploratory investigations and in focused predictive modeling.  Methods for computing the value of information for real-time prediction or for offline planning of additional investments in data collection (*active learning*) can direct dollars and assets to where they will have the highest expected value.  Models can include machinery for effectively computing confidences, allowing for explicit consideration of false positive rates, so as to mediate the invasiveness of actions, and to gauge the impact of policies and investigations on the public.  There is overall great opportunity for better coordinating national technical assets to fuse traditionally disparate data sets into richer models for pursuing answers to questions, as well as monitoring and exploration.  There are opportunities also for enhancing sensory fusion and visualization of sensor data and inferences for commanders and soldiers using our defense systems and platforms, for enhanced situation awareness in peaceful times and during engagements.  And while much of this work involves applying computational tools to classified information, it is possible to develop the underlying approaches and strategies on unclassified data sets and challenge problems.

**Enabling Personalized Education**

Data-centric modeling in education is a challenging area that promises to enhance the way we educate our children.  Several communities of computer scientists have been working to understand how computational models can assist with education.  In particular, efforts in the cognitive science, user modeling, and intelligent tutoring systems communities include building predictive models and decision policies that are used in computer-based tutoring systems and also for developing more ideal teaching strategies.  This research can be viewed as seeking bridges between machine learning and educational psychology.  Other studies to date highlight opportunities for computational methods and analyses to serve as accessories to teachers.  For example, predictive models may one day diagnose students' learning styles and then personalize presentation and content in a way that is engaging and efficient.  Such predictive models have already been constructed from data collected from thousands of K-12 students using computer-based instructional tools in their schools, yielding predictions about which educational material will be most effective for each individual student based on his or her particular errors and understanding of the material.  In addition, predictive models in educational settings can be used to forecast levels of engagement that students have with educational material, and actions that can be taken to enhance engagement or to re-engage students.  An important, related area of work involves generating and operating *simulations* that react intelligently to the actions of individuals or groups of individuals trying to solve a problem in a simulated space, i.e., reacting to gestures, behaviors, or questions, etc., in a way that responds to the individual students, the subject matter, and the pedagogy – thereby shedding light on the various factors and approaches being simulated.  Beyond usage in these systems, data-centric analyses can also advise designs for the best ways to distribute and combine scarce expert pedagogy with always-available online content and tutoring.  Importantly, any research into education data analytics requires deeply rooted involvement of teachers and tutors from the start – as they can not only help steer these analyses from their past experiences but they can also take the new knowledge that is generated and incorporate it into their teaching moving forward.

**Enabling the Smart Grid**

Our nation's electric power grid is an antiquity of networking and distribution that can and should be significantly upgraded based on design principles drawn from modern information networking and communications.  Numerous opportunities for enhancing the reliability and efficiency of the power system will require new overlays and architectures that make possible finer-grained control of distribution and metering.  Fine-grained control of distribution will require new kinds of innovations that provide greater transparency and informational awareness – and that allow for modulation of flows and payments at multiple levels of the power grid.  Prediction of load over time plays a central role in today's power generation industry, where ongoing agreements are made among power suppliers on a daily or hourly manner for transferring large quantities of power among regions of the country.  Decisions are based on forecasts about future demand and availabilities of power from different sources.  These forecasts of future demands are still largely done manually and in a coarse manner.  There are significant opportunities for collecting data on power loads over time, learning predictive models from that data, and employing decision analyses to generate intelligent power distribution strategies.  For example, researchers have shown that instrumenting a home with just three sensors – for electricity, power, and water – makes it possible to determine the resource usage of individual appliances.  There is an opportunity to transform homes and larger residential areas into rich sensor networks, and to integrate information about personal usage patterns with information about energy availability and energy consumption, in support of evidence-based power management.  Directions include developing new systems that provide people with tools to encode preferences about the timing of device usage and interfaces/communication protocols that enable devices to "talk" independently with the grid.  Many rich scenarios for power load balancing and cost optimization are enabled by such tools.  For example, on a hot summer day, HVAC compressors at multiple homes and facilities in a region can be coordinated and sequenced rather than being allowed to operate in a haphazard, independent manner so as to minimize peak loads while delivering effective cooling.  Such systems can gracefully back off during high load situations, according to cooling preferences and willingness to pay, rather than impose large-scale brownouts on a region.  As another example, multiple tasks (e.g., clothes drying) can be deferred to times when power is less expensive, and the decisions about the best timing of usage in advance of a preferred deadline can be supported by predictive models.  We believe that urgent challenges are on the horizon in this realm and these will drive efforts in the near term.  Such challenges include developing approaches to scheduling the charging of large numbers of electric vehicles in a manner that balances the load within neighborhoods, across towns, and across different regions of the larger power grid (as large-scale commercial availability of electric vehicles will start in the autumn of 2010).

**Technical Approaches to Enhancing Data Privacy**

The variety and volume of data collected, and the potential to use this to improve our daily lives, will continue to grow for the foreseeable future.  While the potential benefits are great, for numerous application areas, data privacy and data security must be addressed to achieve the greatest benefits while protecting civil liberties.  These privacy and civil liberties issues are central, and will require a modernization of existing policies for collecting and using data about individuals.  Importantly, there is a key role for technology as well as political process in managing the tradeoff between privacy and the benefits of collecting and using data.  For example, consider the potential for training a predictive model to discover which treatment works best for a new flu strain that is spreading across the country, based on real-time medical records from thousands of emergency rooms and hospitals.  In the past, the only methods to train such a predictive model would have required first combining the data from these thousands of organizations into a central data base, potentially raising privacy concerns.  Today *privacy enhanced* machine learning methods exist that enable training such a model without the need to create a central database.  Instead of centralizing the data, these approaches distribute the machine learning computation, sharing encrypted intermediate results so that the data need never leave the hospital that collected it. Another approach to privacy is *differential privacy,* where special forms of noise are adding to data to obscure peoples' identities without incurring significant losses in the accuracy of inferences.  Other approaches aim to make sharing transparent and controllable, allowing people to both view and make decisions about the data they share.  Providing such views and controls can allow people to make tradeoffs based on their preferences.  For example, people may be comfortable with providing specific types of data in return for enhanced personalization that is enabled by sharing information.  At times, people may wish to share data in an altruistic manner to assist, for example, with healthcare studies or with the crowd-sourcing of traffic flows in a city.  Computational tools can allow people to specify how and when their data can be used, including the type and quantity of data that may be collected.  At the same time, service providers who seek access to personal data can work to understand the varying preferences about privacy and sharing and work to optimize their services based on data that people are willing to share.  To manage the important privacy issues, it will be important to invest in further research into technical approaches to privacy and sharing, to add flexibility as society determines how we can best achieve the benefits of this new data while protecting civil liberties.

**Summary**

Today, data available via the Internet, sensor networks, and new and higher resolution sensors across the sciences allow us to capture more data about people and the world than ever before – and the quantities of data available are accelerating.  Coupled with recent advances in machine learning and reasoning, as well as rapid rises in computing power and storage, we are transforming our ability to make sense of these increasingly large, heterogeneous, noisy or incomplete datasets collected from a variety of sources; to visualize and infer important new knowledge from the data; and to guide action and policies in mission-critical situations, enabling us to make the best decisions. The pipeline of data → predictive models → decision analyses will transform many facets of our daily lives, from healthcare delivery to transportation to energy and the environment. These methods will be critical for Federal agencies tasked with protecting America from threats. And they have the potential to alter how we educate the next generation, how we interact with one another, and how we protect our personal privacy and security in an era of constant connectivity and unfiltered access. In this special series describing how we go *from data to knowledge to action*, we illustrate how data analytics is critical to address our nation's priorities and to ensure our nation's prosperity well into the 21$^{st}$ century – and we provide specific examples of directed investment by mission-driven Federal agencies.